\documentclass{amsart}
\usepackage{amssymb}
\usepackage{amsfonts}
\usepackage{amsmath}
\usepackage{lastpage}
\usepackage[legalpaper,bookmarks=true,colorlinks=true,linkcolor=blue,citecolor=blue]{hyperref}
\usepackage{graphicx}%
\usepackage{fancyhdr}
\usepackage{color}
\usepackage[mathlines]{lineno}
\usepackage{lscape}
\usepackage{epsfig}
\usepackage{natbib}
\usepackage[]{listings}
\usepackage{geometry}
\usepackage{tgbonum}
\fontfamily{qcr}\selectfont

\newtheorem{theorem}{Theorem}
\theoremstyle{plain}

\newtheorem{proposition}{Proposition}

\numberwithin{equation}{section}

\begin{document}
\Large
\title{On some properties of the new Sine-skewed Cardioid Distribution}

\begin{abstract} The new Sine Skewed Cardioid (ssc) distribution been just introduced and characterized by Ahsanullah (2018). Here, we study the asymptotic properties of its tails by determining its extreme value domain, the characteristic function, the moments and likelihood estimators of the two parameters, the asymptotic normality of the moments estimators and the random generation of data from the \textit{ssc} distribution. Finally, we proceed to a simulation study to show the performance of the random generation method and the quality of the moments estimation of the parameters.\\

\bigskip\noindent

\noindent $^{(1)}$ Cherif Mamadou Moctar Traor\'e.\\
Universit\'e des Sciences, des Technique et des Technologies, Bamako, Mali\\
Email : cheriftraore75@yahoo.com\\

\noindent $^{(2)}$ Moumouni Diallo\\
Universit\'e des Scences Economiques et de Gestion (USEG), Bamako, Mali.\\
Email : moudiallo1@gmail.com.\\

\noindent $^{(3)}$ Gane Samb Lo.\\
LERSTAD, Gaston Berger University, Saint-Louis, S\'en\'egal (main affiliation).\newline
LSTA, Pierre and Marie Curie University, Paris VI, France.\newline
AUST - African University of Sciences and Technology, Abuja, Nigeria\\
gane-samb.lo@edu.ugb.sn, gslo@aust.edu.ng, ganesamblo@ganesamblo.net\\
Permanent address : 1178 Evanston Dr NW T3P 0J9,Calgary, Alberta, Canada.\\

\noindent $^{(4)}$ Mouhamad Ahsanullah\\
Department of Management Sciences. Rider University. Lawrenceville, New Jersey, USA\\
Email : ahsan@rider.edu\\

 \noindent $^{(5)}$ Okereke Lois Chinwendu\\
AUST - African University of Sciences and Technology, Abuja, Nigeria\\
Email : lokereke@aust.edu.ng.\\

 \end{abstract}

\maketitle

\section{Introduction} \label{ssc_sec1}

\cite{ashanullah2018} introduced the new distribution with probability distribution function (\textit{pdf})

\begin{equation*}
f(x)=\frac{1}{2\pi }(1+\lambda \sin x)(1+\rho \cos x)1_{\left[ -\pi ,\pi\right] },
\end{equation*}

\bigskip \noindent associated with the parameters $(\lambda ,\rho )\in \left[ -1,1\right] ^{2}$ and named as the sine-skewed cardioid distribution.\\

\noindent In this note, we will state a number properties of that new law. In particular,  we are going asymptotic properties of its tails by determining its extreme value domain, the moments and likelihood estimators of the two parameters, the asymptotic normality of the moments estimators and the random generation of data from the \textit{ssc} distribution. But, before we proceed, we recall for \cite{ashanullah2018} that the cumulative distribution function
(cdf) is

\begin{equation*}
F(x)=\frac{1}{2}+\frac{1}{2\pi }\left( x-\lambda (\cos x+1)+\rho \sin x -\frac{\lambda \rho }{4}(\cos 2x-1)\right), \ x\in \left[-\pi ,\pi \right] .
\end{equation*}

\noindent To prepare studying the upper tail (in a neighborhood of $\pi$) and the lower tail (in a neighborhood of $-\pi$), we may write, respectively, for $x\in \left[-\pi ,\pi \right]$,

\begin{equation}
F(x)=1-\frac{1}{2\pi }\left( (\pi-x)-\lambda (\cos x+1)+\rho \sin (\pi-x) -\frac{\lambda \rho }{4}(\cos 2x-1)\right) \label{cdfUp}
\end{equation}

\noindent and

\begin{equation}
F(x)=\frac{1}{2\pi }\left( (x+\pi)-\lambda (\cos x+1)-\rho \sin (\pi+x) -\frac{\lambda \rho }{4}(\cos 2x-1)\right). \label{cdfLow}
\end{equation}

\bigskip \noindent We will have to deal with some statistical properties. So, we suppose that we have a sequence $X$, $X_1$, $X_2$, $\cdots$, etc. of real-valued random variables defined on the same probability space $(\Omega,\mathcal{A}, \mathbb{P}$ with \textit{cdf} $F$, then supported by $[-\pi,\pi]$. We define the sequence of empirical maxima and the minim

$$
X_{1,n}=\min(X_{1}, \cdots, X_{n}) \ and \ X_{n,n}=\max(X_{1}, \cdots, X_{n}), \ n \geq 1
$$

\bigskip \noindent and the empirical moments

$$
\overline{X}_n=\frac{1}{n} \sum_{1\leq j \leq n} X_j \ ,\ S_n^2=\frac{1}{n-1} \sum_{1\leq j \leq n} \left(X_j-\overline{X}_n\right)^2, \ n \geq 2
$$

\bigskip \noindent and the non-centered second moment

$$
m_{2,n}=\frac{1}{n} \sum_{1\leq j \leq n} X_j^2.
$$

\bigskip \noindent Some graphical illustrations of the \textit{pdf} for some values of the parameters $(\lambda ,\rho )$ are also given by \cite{ashanullah2018} in Figure \ref{fig1}

\begin{figure}
	\centering
		\includegraphics[width=.75\textwidth]{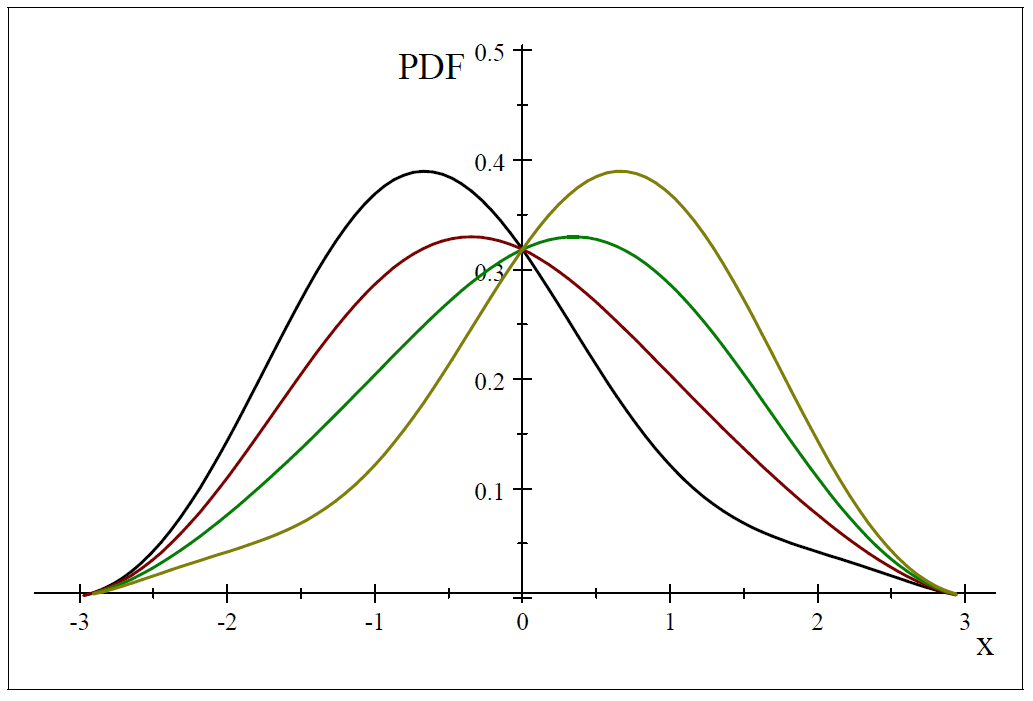}
	\caption{PDF of $f_{sc}(x,1,\lambda ),$Black-$\lambda =-0.6$, Red- $\lambda =-0.2$, Green-$\lambda =0.2$ and Brown-$\lambda =0.6$}
	\label{fig1}
\end{figure}

\bigskip \noindent The rest of the paper is organized as follows. Section \ref{ssc_sec2} is devoted a study of the extreme behavior the the tails of the \textit{ssc} \textit{cdf}. In Section \ref{ssc_sec3}, we deal with the moments and likelihood estimation of the parameters $\lambda$ and $\rho$ and determine the asymptotic laws of the moments estimators. In Section \ref{ssc_sec4}, we provide the characteristic function. Finally, in Section \ref{ssc_sec5}, we propose a method of generating samples from the \textit{ssc} model. Simulation studies are undertaken to test the generation algorithm and next to test the performance of the moments estimators. VB6 Subroutines for the generation methods are given in the appendix. The paper ends by a conclusion section.

\section{Asymptotic Properties of the tails} \label{ssc_sec2}

\begin{theorem} \label{theo_ssc_01}
Define the constants
\begin{eqnarray*}
2\pi \alpha _{1} &=&1+\rho, \ 2\pi \alpha_{2}=\frac{\lambda}{2 }\left(1-\rho\right), \ 2\pi \alpha _{3}=-\frac{\rho}{6} \\
2\pi \alpha _{4} &=&\frac{\lambda}{24} \left(4\rho-1 \right), \ 2\pi \alpha _{5}=\frac{\rho }{120}, \ 2\pi \alpha _{6}=\frac{\lambda}{126} \left( 1-16\rho\right).
\end{eqnarray*}

\bigskip \noindent We have, as $x\rightarrow \pi ,$ 
\begin{eqnarray*}
&&\frac{\alpha _{5}}{\alpha _{6}(\pi -x)}\left( \frac{\alpha _{4}}{\alpha_{5}(\pi -x)}\left( \frac{\alpha _{3}}{\alpha _{4} (\pi -x)}\left( \frac{\alpha _{3}}{\alpha _{2}(\pi -x)}\left( \frac{\alpha _{1}}{\alpha _{2}(\pi-x)}  \right. \right. \right. \right. \\
&&  \left. \left. \left. \left. \left( \frac{1-F(x)}{\alpha _{1}(\pi -x)}\right) -1\right) -1\right),
-1\right) -1\right) -1=O(\pi -x)
\end{eqnarray*}

\bigskip \noindent and as $x\rightarrow -\pi$,
 
\begin{eqnarray*}
&&\frac{\alpha _{5}}{\alpha _{6}(\pi +x)}\left( \frac{\alpha _{4}}{\alpha _{5}(\pi+x)}\left( \frac{\alpha _{3}}{\alpha _{4}(\pi +x)}\left( \frac{\alpha _{3}}{\alpha _{2}(\pi +x)}\left( \frac{\alpha _{1}}{\alpha _{2}(\pi +x)} \right. \right. \right. \right.\\
&& \left. \left. \left. \left. \left( \frac{F(x)}{\alpha _{1}(\pi +x)}\right) -1\right) -1\right) -1\right) -1\right)-1=O(\pi +x).
\end{eqnarray*}
\end{theorem}

\bigskip \bigskip \noindent \textbf{Remark }: Such an expansion is limited to an order 6. But it might be given an any order $k\geq 2.$

\bigskip \noindent \textbf{Proof of Theorem \ref{theo_ssc_01}}. We only give elements for the proof of the first. We use the following elementary expansions

\begin{eqnarray*}
\sin (\pi -x) &=&\sum_{k=0}^{\infty }\frac{(-1)^{k+1}}{(2k+1)!}(\pi-x)^{2k+1} \\
\cos x+1&=&\cos x -\cos \pi =\sum_{k=1}^{\infty }\frac{(-1)^{k+1}}{(2k)!}(\pi -x)^{2k} \\
\cos 2x - 1&=&\cos 2x-\cos 2\pi  =\sum_{k=1}^{\infty }(-1)^{k}\frac{2^{2k}}{(2k)!}(x-\pi)^{2k}.
\end{eqnarray*}

\bigskip \noindent We consider the limited expansion at order 6 at $\pi$ to have

\begin{eqnarray*}
1-F(x) &=&\alpha _{1}(\pi -x)+ \alpha _{2}(\pi -x)^{2}+ \alpha _{3}(\pi -x)^{3}+2\pi \alpha_{4}(\pi -x)^{4} \\
&& +\alpha _{5}(\pi -x)^{5}+ \alpha _{6}(\pi -x)^{6}+O((\pi-x)^{7}.
\end{eqnarray*}

\bigskip \noindent This justifies the expansion at $+\pi$. The situation is the same at $-\pi$ from Formula \label{cdfLow}. The reason is that $+\pi$ and $-\pi$ play the same roles in the developments and the two terms $-\rho \sin (\pi+\theta)$ and $\rho \sin (\pi-\theta)$ are expanded in the same way with respect to $\pi-x$ and $\pi+x$ respectively. $\square$\\

\bigskip \noindent From Theorem \ref{theo_ssc_01}, we directly get the extreme law of $X$ and $Y=1/(\pi +X)$.

\begin{theorem} \label{theo_ssc_02} We have the following properties

\bigskip \noindent (a) $F^{\star }(x)=F(\pi -\frac{1}{x}),x>0$ belong to the Frechet extreme value Domain $D(H_{1})$ that is 
\begin{equation*}
\forall \gamma >0,\text{ }\lim_{x\rightarrow +\infty }\frac{1-F^{\star
}(\gamma x)}{1-F^{\star }(x)}=\gamma ^{-1}
\end{equation*}

\bigskip \noindent and the second order condition 

\begin{equation*}
\forall \gamma >0,\text{ }\lim_{x\rightarrow +\infty }\frac{1}{S(x)}\left( 
\frac{1-F^{\star }(\gamma x)}{1-F^{\star }(x)}-\gamma ^{-1}\right) = 0
\end{equation*}

\noindent for  $S(x)=\alpha_2 x/\alpha_1$, $x>0$.\\

\noindent b) As a consequence $F$ is in the Weibull extreme value Domain $D(H_{-1})$ and we have

$$
n(1+\rho)\left(X_{n,n}-\pi\right) \rightsquigarrow H_{-1}, \ as \ n\rightarrow \infty.
$$

\noindent (c) To have the extreme lower law is found by using $X_{1,n}=-(-X)_{n,n}$ and $(-X)_{n,n}$ and $X_{n,n}$ have the same limit in type, that is

$$
n(1+\rho)\left((-X)_{n,n}-\pi\right) \rightsquigarrow H_{-1}, \ as \ n\rightarrow \infty.
$$
\end{theorem}

\bigskip \noindent \textbf{Proof of Theorem \ref{theo_ssc_02}}. Point (a) is a direct consequence of Theorem \label{theo_ssc_01} at the first order. By Theorem 8 in \cite{ips-wciia-ang}, we have that
 $F \in D(H_1)$ if and only if $F^\ast \in D(H_{-1})$. So (b) holds from (a). Furthermore, by  Proposition 8 in \cite{ips-wciia-ang}, we also have 

\begin{equation}
\frac{X_{n,n}-uep(F)}{uep(F)-F^{-1}(1-1/n))} \rightsquigarrow H_{-1}, \ as \ n\rightarrow \infty. \label{weibullSsc}
\end{equation}

\noindent Now from the expansions in Theorem \label{theo_ssc_01}, we have for any $-1<u<1$
\begin{eqnarray*}
&&1-F(x)=u\\
&\Leftrightarrow&\alpha_1 (\pi-x)+ \alpha_2 (\pi-x)^2 +...+ \alpha_6 (\pi-x)^6+O((\pi-x)^7).
\end{eqnarray*}

\bigskip \noindent We get that, as $x\rightarrow +\pi$, 
$$
(\pi-x) \sim u/\alpha_1,
$$

\bigskip \noindent and

$$
x=F^{-1}(1-u)=\pi-\frac{1}{\alpha_1}u + \frac{\alpha_2}{\alpha_1^3}u^2(1+\varepsilon_2(1))+ \cdots + \frac{\alpha_6}{\alpha_1^7}u^6(1+\varepsilon_6(1))+O(u^7),
$$

\noindent where the function $\varepsilon_h(u)$ go to zero as $u \downarrow 0$. In particular, we have

$$
\pi-F^{-1}(1-u) \sim \frac{1}{\alpha_1}u=\frac{u}{1+\rho}
$$

\bigskip \noindent and, as $n \rightarrow +\infty$

$$
\pi-F^{-1}(1-1/n) \sim \frac{1}{\alpha_1}=\frac{1}{n(1+\rho)},
$$

\bigskip \noindent which combined with Formula \ref{weibullSsc} concludes Point (b) of the proof.\\

\noindent Point (c) is based the \textit{cdf} of $-X$, which is $F^{\bot}(x)=1-F(-x)$, for $-\pi \leq x \leq \pi$. And the expansion of $1-F^{\bot}(x)$ gives the same expansion as in Formula
\label{cdfUp} at $+\pi$. We get the same conclusion as for $X$. $\blacksquare$\\

\noindent \textbf{Interesting Pedagogical Example}. As we can find in \cite{ips-wciia-ang}, page 133, a criterion of belonging of $F$ to $D(H_1)$, when $uep(F)=\pi$ is that $F$ admits a derivative in a right neighborhood of $\pi$ and that

\begin{equation}
\lim_{x \rightarrow \pi} \frac{(\pi-x)F^{\prime}(x)}{1-F(x)}=1. \label{domWeilC}
\end{equation}

\noindent It is also stated that the condition \ref{domWeilC} holds if $F \in D(H_1)$ and $F^{\prime}$ is ultimately non-increasing as $x \nearrow \pi$. Here, the limit holds for all values of $\lambda$ and $\rho$ in $]-1,1[$ although $F^{\prime}$ is ultimately not non-increasing for some values of $\lambda$ and $\rho$ for example, for $\lambda=-0.8$ and 
$\rho=-0.4$, for which $F^{\prime}$ is increasing as shown in Figure \ref{fig1}

\begin{figure}
	\centering
		\includegraphics[width=0.50\textwidth]{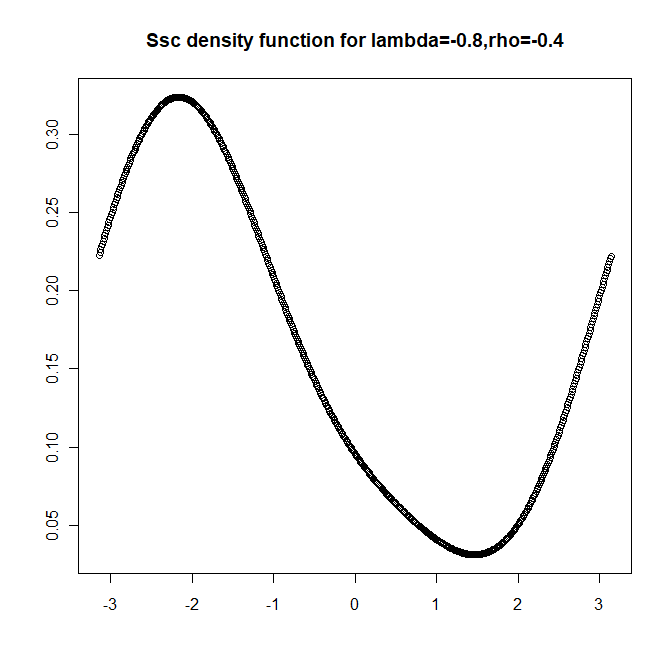}
	\caption{Ssc pdf for $\lambda=-0.8$ and $\rho=-0.4$}
	\label{fig1}
\end{figure}

\bigskip \noindent With theses values, if the practitioner concludes that $F \in D(H_1)$, he is wrong in the use of the rule but his conclusion is correct by coincidence. So it is important to check the ultimate decreasingness of $F^{\prime}$, which is used in the proof of the rule.\\

\section{Parameters estimation} \label{ssc_sec3}

\subsection{Moments estimation}$ $\\

\noindent \textbf{A. Point Estimation}.\\

\noindent The $kth$ moment of $X$ is given by

$$
\mathbb{E}X^k=\frac{1}{2 \pi}\int_{-\pi}^{\pi} x^k(1+\lambda \sin x + \rho \cos x + (\lambda \rho/2) \sin 2x) \ dx.
$$

\bigskip \noindent From the following facts

\begin{eqnarray*}
&& \int_{-\pi}^{\pi} x \cos x \ dx=0, \ \ \int_{-\pi}^{\pi} x \sin x \ dx=2\pi , \ \ \int_{-\pi}^{\pi} x \cos x  \sin x \ dx=-\pi/2\\
&&\int_{-\pi}^{\pi} x^2 \cos x \ dx=-4\pi, \  \int_{-\pi}^{\pi} x^2 \sin x \ dx=0, \ \  \int_{-\pi}^{\pi} x^2 \sin x\cos x \ dx=0,
\end{eqnarray*}

\bigskip \noindent we get

$$
m=\mathbb{E}X=\lambda (1-\rho/8), \ m_2=\mathbb{E}X^2=\pi^2/3-2 \rho.
$$

\noindent The moments estimators are solutions of the system of two equations : $m=\overline{X}_n$ and $m_2=m_{2,n}$. We immediately have, for $n\geq 1$,

\begin{equation}
\hat{\rho}_n=(\pi^2/3-m_{2,n})/2 \ and \ \hat{\lambda}_n=\frac{4\overline{X}_n}{4-\hat{\rho}_n}=\frac{8\overline{X}_n}{8-\pi^2/3+m_{2,n}}.
\end{equation}

\bigskip \noindent We may need to check such estimation by a simulation study. We will do this in Section \ref{ssc_sec4} where we propose a simple method for generating data for the \textsl{ssc} distribution. For now we want to do more on the moment problem. In the appendix, we study the integrals, for $n\geq 0$,

$$
I_n=\int_{-\pi}^{\pi} x^n \cos x \ dx, \ J_n=\int_{-\pi}^{\pi} x^n \sin x \ dx, \  and \ H_n=\int_{-\pi}^{\pi} x^n \sin 2x \ dx.\\
$$

\bigskip \noindent We established the following recurrence formula

\begin{equation}
I_0=0, \forall n\geq 2, I_{2n}=-4n \pi^{2n-1}- 2n(2n-1) I_{2(n-1)} \ and \ \forall n\geq 0, I_{2n+1}=0, \label{recIP},
\end{equation}

\begin{equation}
\forall n\geq 0, J_{2n}=0 \ and \ \forall n\geq 1, J_{2n+1}=2\pi^{2n+1} - 2n(2n+1) J_{2(n-1)+1}, \label{recJP}
\end{equation}

\bigskip \noindent and

\begin{equation}
\forall n\geq 0, H_{2n}=0 \ and \ \forall n\geq 1, H_{2n+1}=-\frac{\pi^{2n+1}}{2} - \frac{n(2n+1) H_{2(n-1)+1}}{2}. \label{recHP}
\end{equation}

\bigskip \noindent and for the needs of that paper, we have computed

\begin{eqnarray*}
&& I_1=0, \ I_2=(2\pi) (-2\pi), \ I_3=0, \ I_4=4(2\pi)(6-\pi^2))\\
&& J_1=2\pi, \ J_2=0, \ J_3=(2\pi)(\pi^2-6), \ J_4=0\\
&& H_1=(2\pi) (-1/4), \ H_2=0, H_3=(2\pi)((3-2\pi^2)/8), \ H_4=0.
\end{eqnarray*}

\noindent With such facts, we easily have

\begin{eqnarray}
\mathbb{E}X^3&=&\lambda (\pi^2-6)+\frac{\lambda \rho}{8}(3-2\pi^2) \notag \\
\mathbb{E}X^4&=&\frac{\pi^4}{5}-4\rho(\pi^2-6). \label{moment34}
\end{eqnarray}

\bigskip \noindent We will see how we need these parameters for the asymptotic laws of the moment estimators.

\section{The characteristic function} \label{ssc_sec4}

\begin{proposition} The characteristic function (\textit{fc}) of $X$ is given, for $t \notin \{-2,-1,0,1,2\}$, bt
$$
\psi(t)=\frac{1}{\pi}\left(\frac{1}{t}+\frac{\lambda}{i(t^2-1)}-\frac{\rho t}{t^2-1}- \frac{\lambda \rho}{i(t^2-4)} \right) sin(\pi t)
$$

\noindent and is extended to values in $\{-2,-1,0,1,2\}$ by continuity of the  \textit{fc}.
\end{proposition}

\bigskip \noindent \textbf{Proof}. We may write for $t \in \mathbb{R}$ and $x \in [-\pi, \pi]$,

\begin{eqnarray*}
e^{it}f(x)&=&\frac{1}{2\pi} e^{itx} \left[1 +\frac{\lambda}{2i}\left(e^{ix}-e^{-ix}\right)+\frac{\rho}{2}\left(e^{ix}+e^{-ix}\right)+\frac{\lambda \rho}{4}\left(e^{i2x}-e^{-i2x}\right) \right]\\
&=&\frac{1}{2\pi}  \left[e^{itx} +\frac{\lambda}{2i}\left(e^{ix(t+1)}-e^{ix(t-1)}\right)+\frac{\rho}{2}\left(e^{ix(t+1)}+e^{ix(t-1)}\right) \right.\\
&+& \left. \frac{\lambda \rho}{4}\left(e^{ix(t+2)}-e^{-ix(t-2)}\right) \right].
\end{eqnarray*}

\noindent Integrating from $-\pi$ to $+\pi$ leads to the announced results. $\square$\\

\noindent \textbf{B. Asymptotic Normality of the moment estimators and Statistical tests}.\\

\noindent The moments estimators are treated by using the function empirical process defined, for any $n\geq 1$ and $h \in L^2(\mathbb{P}_X)$, by

$$
\mathbb{G}_n(h)=\frac{1}{\sqrt{n}} \sum_{j=1}^{n} (h(X_j)-\mathbb{E}(h(X_j))),
$$

\noindent as explained in \cite{gsloptlms} and \cite{ips-wciiia-ang}. For a more general source, the book by \cite{vaart} is one the best in the field but we no need the great artillery provided there. We have the following result.

\begin{theorem} we have the asymptotic laws of the moment estimators, as $n\rightarrow +\infty$,

$$
\left(\sqrt{n}(\hat{\rho}_n-\rho), \sqrt{n}(\hat{\lambda}_n-\lambda)\right) \rightsquigarrow \mathcal{N}_2\left(0, \Sigma\right),
$$

\noindent with 

$$
\Sigma_{11}=\frac{\mathbb{V}ar(X^2)}{4}, \ \  \Sigma_{11}=\frac{\mathbb{V}ar(X+\lambda X^2)}{\mu^2}
$$

\noindent and

$$
\Sigma_{12}=\Sigma_{21}=-\frac{\mathbb{C}ov(X^2, X+\lambda X^2)}{2\mu}.
$$
\end{theorem}

\bigskip \noindent \textbf{Remark}. As anticipated in Formula \ref{moment34}, the asymptotic laws need the first four moments of $X$.\\

\noindent \textbf{Proof}.  Let us set $h_{\ell}(x)=x^{\ell}$ for $x \in \mathbb{R}$. By applying the methods in \cite{gsloptlms}, we get

$$
\sqrt{n}(\hat{\rho}_n-\rho)=\mathbb{G}_n(-h_2/2) +o_{\mathbb{P}(1)}.
$$

\bigskip \noindent Next, we have

$$
\hat{\lambda}_n=\frac{8\left(m + \frac{1}{\sqrt{n}} \mathbb{G}_n(h_1)\right)}{8-\pi^2/3+m_2 + \frac{1}{\sqrt{n}} \mathbb{G}_n(h_2)}.
$$

\bigskip \noindent By Lemma 2 in \cite{gsloptlms}, we get that for $\mu=8-\pi^2/3+m_2$,

$$
\sqrt{n}(\hat{\lambda}_n-\lambda)=\mathbb{G}_n((h_1+\lambda h_2)/\mu) +o_{\mathbb{P}(1)}. 
$$

\noindent By using the functional Brownian stochastic process $\mathbb{G}$, which is the weak limit of $\mathbb{G}_n$ and defined by the variance-covariance function

$$
\Gamma(h,k)=\int_{\mathbb{R}} (h(x)-\mathbb{E}h(X))(k(x)-\mathbb{E}k(X))d\mathbb{P}_X(x),
$$

\noindent where $(h,k) \in L^2(\mathbb{P}_X)^2$, we get that

$$
\left(\sqrt{n}(\hat{\rho}_n-\rho), \sqrt{n}(\hat{\lambda}_n-\lambda)\right) \rightsquigarrow \mathcal{N}_2\left(0, \Sigma\right),
$$

\noindent with 

$$
\Sigma_{11}=\frac{\mathbb{V}ar(h_2(X^2)}{4}, \ \  \Sigma_{11}=\frac{\mathbb{V}ar(h_1(X)+\lambda h_2(X)}{\mu^2}
$$

\noindent and

$$
\Sigma_{12}=\Sigma_{21}=-\frac{\mathbb{C}ov(h_2(X), h_1(X)+\lambda h_2(X))}{2\mu}. \ \blacksquare 
$$

\subsection{Maximum Likelihood Estimators}

\noindent we are going to see that the \textit{ML}-estimators are not defined here. We begin the remark that the first three linear differential operators in $(\lambda,\rho)$ of $f$ are 

\begin{eqnarray*}
&&2\pi Df(u,v)=\sin x (1+\rho \cos x) u + \cos x (1+\lambda \sin x) v,\\
&&2\pi D^2f(u,v)=(\cos x \sin x) uv, \  (u,v) \in \mathbb{R},\\
&&D^3 f(u,v)=0.
\end{eqnarray*} 

\noindent By applying the Taylor-Lagrange-Cauchy formula (see \cite{valiron}, page 233) :  for $(\lambda,\lambda_0,\rho,\rho_0) \in [-1,1]^4$, for $|\theta_i|<1$, $i=1,2$

\begin{eqnarray*}
f(x,\lambda,\rho)&=&f(x,\lambda_0,\rho_0)+ Df(\lambda-\lambda_0,\rho-\rho_0) +\frac{1}{2} D^2f(\lambda-\lambda_0,\rho-\rho_0)\\
&+&D^2f(\theta_1(\lambda-\lambda_0),\theta_2(\rho-\rho_0)),
\end{eqnarray*}

\noindent and we get

\begin{eqnarray*}
f(x,\lambda,\rho)&=&f(x,\lambda_0,\rho_0)+ \sin x (1+\rho \cos x) (\lambda-\lambda_0)+ \cos x (1+\lambda \sin x) (\rho-\rho_0)\\
&+& (\cos x \sin x) (\lambda-\lambda_0)(\rho-\rho_0)
\end{eqnarray*}

\noindent First, for $x \notin \{\mp\pi, \pm \pi/2\}$, the zeros of $f$ are $-1/\sin x$ and $-1/\cos x$ which do not belong to $[-1,1]$. Even on $\mathbb{R}$, not requiring that $f$ be non-negative to make it a probability density function, Formula which becomes at any critical point $(\lambda_0,\rho_0)$ of this $C^1$-function (in $\lambda$ and in $\rho$),

\begin{eqnarray*}
f(x,\lambda,\rho)&=&(\cos x \sin x) (\lambda-\lambda_0)(\rho-\rho_0).
\end{eqnarray*}

\noindent So, for a fixed $x \notin \{\mp\pi, \pm \pi/2\}$, there can be an extremum point $(\lambda_0,\rho_0)$ for the likelihood function.\\

\section{Generation} \label{ssc_sec5}

\noindent Since the \textit{cdf} $F$ is explicitly known and is strictly increasing and continuous, we may use the dichotomous algorithm to find the inverse of $F$. It works as follows. Given 
$v\in [0,1]$, to find $x$ such that $F(x)=v$, we fix the number of decimals of the solution $x$ denoted \textit{nbrDec}. In the Appendix, beginning by page \pageref{ssc_sample_generations}, the VB 6 code for computing the \textit{cdf} $F$ is given in page \pageref{ssc_cdf}, the Dichotomous algorithm is described in page \pageref{dichotmous_algo} and implemented in VB6 in \pageref{dichotmous_implement}. Finally, the VB 6 subroutine which generates a sample of an arbitrary size is given in  \pageref{ssc_random}.\\

\noindent \textbf{A. Numerical test of the computer programs}.\\

\noindent For different values for $\lambda$ and $\rho$ in $]-1,1[$, samples of size $n=1000$ are generated and the comparison between the exact means and the second moments as given in Formua are compared with the sample counterparts. Table \ref{tab1} demonstrates the quality of the generation.

\begin{table}
	\centering
		\begin{tabular}{lllllll}
		\hline  \hline 
	$\lambda$ & $\rho$ & mean (E) & mean (M)& 2nd moment (E) & 2nd moment (M) & quotient (Q)\\	
		\hline  \hline 
0.9&-0.9 	&1.1025 	& 1.2322 & 5.0898 & 5.0281 & 1.0122\\
0.9&-0.6 	&1.035 		& 1.0641 & 4.4898 & 4.4796 & 1.0022\\
0.9&-0.3 	&0.9675 	& 0.9396 & 3.8898 & 3.8987 & 0.9977\\
0.9&0.1 	&0.8775 	& 0.8796 & 3.0898 & 3.2221 & 0.9589\\
0.9&0.4 	&0.81 		& 0.8048 & 2.4898 & 2.5953 & 0.9593\\
0.9&0.7 	&0.7425 	& 0.7536 & 1.8898 & 1.8690 & 1.0111\\
0.9&0.9 	&0.6975 	& 0.7077 & 1.4898 & 1.4810 & 1.0059\\
-0.9&0.9 	&-0.6975 	&-0.6972 & 1.4898 & 1.5574 & 0.9566\\
-0.9&0.7 	&-0.7425 	&-0.7773 & 1.8898 & 1.8280 & 1.0338\\
-0.9&0.4 	&-0.81 		&-0.8218 & 2.4898 & 2.2965 & 1.0841\\
-0.9&0.1 	&-0.8775 	&-0.9273 & 3.0898 & 3.2964 & 0.9373\\
-0.9&-0.3 &-0.9675 	&-1.0273 & 3.8898 & 3.9613 & 0.9819\\
-0.9&-0.6 &-1.035 	&-0.9654 & 4.4898 & 4.4845 & 1.0011\\
-0.9&-0.9 &-1.1025 	&-1.1300 & 5.0898 & 5.1902 & 0.9806\\
\hline 
		\end{tabular}
	\caption{Legend : (E) : Exact, (M) empirical, (Q) : quotient exact second moment to empirical second moment}
	\label{tab1}
\end{table}

\bigskip \noindent \textbf{B. Moment estimation}.\\

\noindent Based on the generation techniques as introduced above, we also report interesting performances for the estimation of the parameter when the sizes varies in Table \ref{tab2}. The Mean Absolute Error (MAE) and the  Mean Square-root Quadratic Error (MSQE) are reported both for $\lambda$ and $\rho$. These simulations have been for $\lambda=\rho=-0.9$.

\begin{table}
	\centering
\begin{tabular}{llllllllll}
\hline \hline
Error n 				&10	&50	&100	&200	&300	&400	&500	&750	&1000\\
\hline \hline
MAE-$\lambda$		&41.17 	&18.44 	&11.88 	&10.41	&7.39 	&5.93 	&5.43 	&5.08 	&3.91\\
SMQE-$\lambda$	&51.97	&22.49	&15.04	&12.07	&9.81		&7.74		&6.78		&6.17		&5.20\\
MAE-$\rho$			&29.83	&15.20	&10.05	&8.51		&4.97		&5.56		&4.59		&3.56		&3.14\\
SMQE-$\rho$			&37.44	&18.10	&13.10	&10.09	&6.49		&7.17		&5.66		&4.41		&2.89\\
\hline \hline
\end{tabular}
	\caption{Errors given in $100$ multiples}
	\label{tab2}
\end{table}

\begin{figure}
	\centering
		\includegraphics[width=1.00\textwidth]{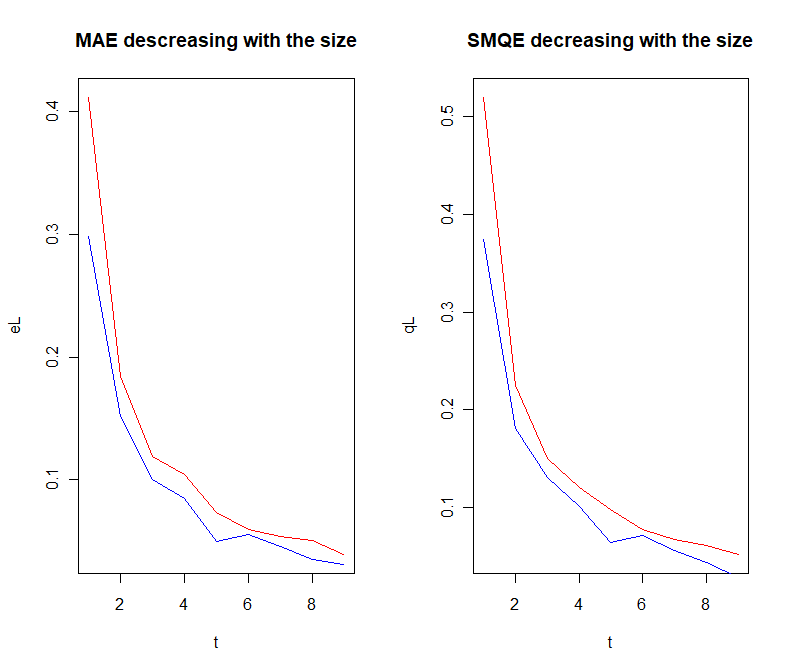}
	\caption{\textcolor{red}{Red} : related to $\rho$. \textcolor{blue}{Blue} : errors related to $\rho$}
	\label{figSim}
\end{figure}

\noindent  Figure \ref{figSim} shows how both errors decreases to zero.

\section{Conclusion} This is an immediate contribution of the study of the Sine Skewed Caedioid distribution. Further deeper properties will be addressed later.

\newpage
\noindent \textbf{Appendix}.\\

\noindent \textbf{I. Integral Computations}. \label{ssc_integral_computations}\\

\noindent Let us define, for $n\geq 0$,

$$
I_n=\int_{-\pi}^{\pi} x^n \cos x \ dx, \ \int_{-\pi}^{\pi} x^n \sin x \ dx, \ \int_{-\pi}^{\pi} x^n \cos 2x \ dx \ and \ \int_{-\pi}^{\pi} x^n \sin 2x \ dx.\\
$$

\noindent \textbf{A - Computation of $I_n$}. For $n=0$, 

$$
I_0=\int_{-\pi}^{\pi} \cos x \ dx=\int_{-\pi}^{\pi} d(\sin x)=0
$$ 

\noindent and for $n=1$,

$$
I_1=\int_{-\pi}^{\pi} x d(\sin x)=\lbrack x \sin x \rbrack^{-\pi}_{\pi} - \int_{-\pi}^{\pi} -d(\cos x)=0
$$ 

\noindent and for $n=2$

\begin{eqnarray*}
I_2&=&\int_{-\pi}^{\pi} x^2 d(\sin x)=\rbrack x^2 \sin x \lbrack^{-\pi}_{\pi}- 2 \int_{-\pi}^{\pi} -2xd(\cos x)\\
&=& \left( 4 \lbrack x \cos x \rbrack^{-\pi}_{\pi}-\int_{-\pi}^{\pi} \cos x) \right)\\
&=& 4 \lbrack x \cos x \rbrack^{-\pi}_{\pi}=-4\pi.
\end{eqnarray*}

\noindent Now, the general guess is that $I_{2n+1}=0$ for $n\geq 0$. Since this already holds $n=0$, let us remark that for $n\geq 1$,

\begin{eqnarray*}
I_{2n+1}&=& \int_{-\pi}^{\pi} x^{2n+1} d(\sin x)=\rbrack x^{2n+1} \sin x \lbrack^{-\pi}_{\pi}- (2n+1) \int_{-\pi}^{\pi} -x^{2n} d(\cos x)\\
&=& (2n+1) \int_{-\pi}^{\pi} x^{2n} d(\cos x)= (2n+1)\lbrack x^{2n} \cos x \rbrack^{-\pi}_{\pi} - (2n)(2n+1) \int_{-\pi}^{\pi} -x^{2n-1} \cos x \ dx\\
&=&-(2n)(2n+1)I_{2(n-1)+1}.
\end{eqnarray*}

\noindent By an descendent induction, we will have $I_{2n+1}=C_n I_1=0$. Now, we have to find $I_{2n}$, for $n\geq 1$, which is

\begin{eqnarray*}
I_{2n}&=& \int_{-\pi}^{\pi} x^{2n} d(\sin x)=\rbrack x^{2n} \sin x \lbrack^{-\pi}_{\pi} - (2n) \int_{-\pi}^{\pi} (-x^{2n-1}) d(\cos x)\\
&=& (2n) \int_{-\pi}^{\pi} x^{2n-1} d(\cos x)= \lbrack x^{2n-1} \cos x \rbrack^{-\pi}_{\pi} - (2n)(2n-1) \int_{-\pi}^{\pi} -x^{2(n-1)} \cos x \ dx\\
&=&-4n \pi^{2n-1}- 2n(2n-1) I_{2(n-1)}.
\end{eqnarray*}

\noindent For example, we find again $I_2=-4\pi$ for $n=1$.\\

\noindent \textbf{B - Computation of $J_n$}. For $n=0$, we have

$$
J_0=\int_{-\pi}^{\pi} \sin x \ dx=\int_{-\pi}^{\pi} d(-\cos x)=0
$$ 

\noindent and for $n=1$,

$$
J_1=\int_{-\pi}^{\pi} x d(-\cos x)=- \lbrack x \cos x \rbrack^{-\pi}_{\pi} + \int_{-\pi}^{\pi} -d(\sin x)=2\pi.
$$

\noindent Our guess is that $J_{2n}=0$ for $n\geq 0$ and we have for $n\geq 1$,

\begin{eqnarray*}
J_{2n}&=& \int_{-\pi}^{\pi} x^{2n} d(-\cos x)=\rbrack x^{2n} \cos x \lbrack^{-\pi}_{\pi}+ (2n) \int_{-\pi}^{\pi} x^{2n-1} d(\sin x)\\
&=& (2n) \int_{-\pi}^{\pi} x^{2n-1} d(\sin x)= (2n) \lbrack x^{2n-1} \sin x \rbrack^{-\pi}_{\pi} - (2n)(2n-1) \int_{-\pi}^{\pi} x^{2(n-1)} \sin x \ dx\\
&=&-(2n)(2n-1)J_{(n-1)}.
\end{eqnarray*}

\noindent By an descendent induction, we will have $J_{2n}=C_n J_0=0$. Now, he have to find $J_{2n+1}$, for $n\geq 1$, which is

\begin{eqnarray*}
J_{2n+1}&=& \int_{-\pi}^{\pi} x^{2n+1} d(-\cos x)=-\rbrack x^{2n+1} \cos x \lbrack^{-\pi}_{\pi}+ (2n+1) \int_{-\pi}^{\pi} x^{2n} d(\sin x)\\
&=& 2 \pi^{2n+1} +(2n+1) \lbrack x^{2n} \sin x \rbrack^{-\pi}_{\pi} - (2n)(2n+1) \int_{-\pi}^{\pi} x^{2(n-1)+1} \sin x \ dx\\
&=&2\pi^{2n+1} - 2n(2n+1) J_{2(n-1)+1}.
\end{eqnarray*}

\noindent \textbf{C - Computation of $H_n$}. For $n=0$, we have

$$
H_0=\int_{-\pi}^{\pi} \sin x d(\sin x)=(1/2) \int_{-\pi}^{\pi} d(\sin^2 x)=0,
$$ 

\noindent and for $n=1$,

\begin{eqnarray*}
H_1&=&(1/2)\int_{-\pi}^{\pi} x \sin 2x \ dx=- (1/4)\int_{-\pi}^{\pi} x d(\cos 2x)\\
&=&-(1/4) \left(\lbrack x \cos 2x \rbrack^{-\pi}_{\pi} - \int_{-\pi}^{\pi} (1/2) d(\sin 2x)\right)\\
&=&-\pi/2.
\end{eqnarray*}

\noindent We still guess that $H_{2n}=0$ for $n\geq 0$ and find, for $n\geq 0$,

\begin{eqnarray*}
H_{2n}&=&(1/2)\int_{-\pi}^{\pi} x^{2n} \sin 2x \ dx=- (1/4)\int_{-\pi}^{\pi} x^{2n} d(\cos 2x)\\
&=&-(1/4) \left(\lbrack x^{2n} \cos 2x \rbrack^{-\pi}_{\pi} - (1/2) (2n) \int_{-\pi}^{\pi} x^{2n-1} d(\sin 2x)\right)\\
&=&(1/4)n \int_{-\pi}^{\pi} x^{2n-1} d(\sin 2x)=(1/4)n \left( \lbrack x^{2n-1} \sin 2x \rbrack^{-\pi}_{\pi} - (2n-1) \int_{-\pi}^{\pi} x^{2(n-1)} \sin 2x \ dx\right)\\
&=&-(1/4)n(2n-1)H_{2(n-1)}, 
\end{eqnarray*}

\noindent and we conclude that  $H_{2n}=0$ for all $n\geq 0$, based on that $H_0=0$ and the unveiled descendent induction formula above. Nor for all $n\geq 0$,

\begin{eqnarray*}
H_{2n+1}&=&(1/2)\int_{-\pi}^{\pi} x^{2n+1} \sin 2x \ dx=- (1/4)\int_{-\pi}^{\pi} x^{2n+1} d(\cos 2x)\\
&=&-(1/4) \left(\lbrack x^{2n+1} \cos 2x \rbrack^{-\pi}_{\pi} - (1/2) (2n+1) \int_{-\pi}^{\pi} x^{2n} d(\sin 2x)\right)\\
&=&-(1/4)\left( (2\pi^{2n+1}) - (1/2) (2n+1) \left(\lbrack x^{2n} \sin 2x \rbrack^{-\pi}_{\pi} - (2n) \int_{-\pi}^{\pi} x^{2n-1} \sin 2x) \right)\right)\\
&=&-(1/2)\pi^{2n+1} - (1/4)(1/2) (2n+1)(2n) \int_{-\pi}^{\pi} x^{2n-1} \sin 2x)\\
&=&-(1/2)\pi^{2n+1} - (1/2)n(2n+1) H_{2(n-1)+1}
\end{eqnarray*}

\noindent We conclude

\begin{equation}
I_0=0, \forall n\geq 2, I_{2n}=-4n \pi^{2n-1}- 2n(2n-1) I_{2(n-1)} \ and \ \forall n\geq 0, I_{2n+1}=0, \label{recI},
\end{equation}

\begin{equation}
\forall n\geq 0, J_{2n}=0 \ and \ \forall n\geq 1, J_{2n+1}=2\pi^{2n+1} - 2n(2n+1) J_{2(n-1)+1}, \label{recJ}
\end{equation}

\noindent and

\begin{equation}
\forall n\geq 0, H_{2n}=0 \ and \ \forall n\geq 1, H_{2n+1}=-\frac{\pi^{2n+1}}{2} - \frac{n(2n+1) H_{2(n-1)+1}}{2}. \label{recH}
\end{equation}

\bigskip \noindent For the needs of this paper, we have

\begin{eqnarray*}
&& I_1=0, \ I_2=(2\pi) (-2\pi), \ I_3=0, \ I_4=4(2\pi)(6-\pi^2),\\
&& \ I_5=0, \ I_6=-6(2\pi)(\pi^4-20\pi^2+120), \ I_7=0, \ I_8=-8(2\pi)(\pi^6-42\pi^4+840\pi^2-5040) \\
&& J_1=2\pi, \ J_2=0, \ J_3=(2\pi)(\pi^2-6), \ J_4=0,\\
&& J_5=2\pi(\pi^4-20\pi^2+120), \ J_6=0, \ J_7=2\pi(\pi^6-42\pi^4+840\pi^2-5040), \ J_8=0,\\
&& H_1=(2\pi) (-1/4), \ H_2=0, H_3=(2\pi)((3-2\pi^2)/8),\ H_0=0, \\
&& H_5=-2\pi\frac{(2\pi^4-10\pi^2+15)}{8}, \ H_6=0, \ H_7=-2\pi\frac{(4\pi^6-42\pi^4+210\pi^2-315)}{16} .
\end{eqnarray*}

\begin{eqnarray*}
\mathbb{E}X^5&=&\lambda (\pi^4-20\pi^2+120)-\frac{\lambda \rho}{16}(2\pi^4-10\pi^2+15) \notag \\
\mathbb{E}X^6&=& \frac{\pi^6}{7}-6\rho(\pi^4-20\pi^2+120) \notag \\
\mathbb{E}X^7&=&\lambda (\pi^6-42\pi^4+840\pi^2-5040)-\frac{\lambda \rho}{32}(4\pi^6-42\pi^4+210\pi^2 +315) \notag \\
\mathbb{E}X^8&=&\frac{\pi^8}{9}-8\rho(\pi^6-42\pi^4+840\pi^2-5040). \label{moment35}
\end{eqnarray*}



\newpage
\noindent \textbf{II. Samples generations}. \label{ssc_sample_generations}\\

\normalsize
\noindent \textbf{Description of the dichotomous algorithm}. \label{dichotmous_algo}
\begin{lstlisting}
0. Fix the count index of the decimals countDec to -1 : countDec=-1.

1. Fix x=1 and compute F(1).

2. IF F(1)=v. Take x=1. Stop

3. IF F(1)<v, then

		3a. Fix x=1 and set h=1.
		3b. Progressively increment x by h and test 
				and compare F(x) and v.
				3bA. IF F(x)=v. x is the searched number.Stop
				3bB. IF F(x)<v, Goto 3bA and continue 
						 the incrementation.
				3bC. IF F(x)>v, then
												Decrement by h : x=x-h
												Increment countDec
												If countDec<nbrDec
														take h=h/10
														Goto 3b
												If countDec=nbrDec. Stop

3. IF F(1)>v, then

		4a. Fix x=1 and set h=0.1
		4b. Progressively decrement x by h : x=x-h and test and compare 
				F(x) and v.
				4bA. IF F(x)=v. x is the searched number.Stop
				4bB. IF F(x)>v, Goto 4bA and continue 
						 the decrementation
				4bC. IF F(x)<v, then
												increment by h : x=x-h
												Increment countDec
												If countDec<nbrDec-1
														take h=h/10
														Goto 4b
												If countDec=nbrDec. Stop
\end{lstlisting}

\noindent \textbf{Implementation of the dichotomous algorithm in Visual Basic 6}.\label{dichotmous_implement} 
\begin{lstlisting}
Function sscinv(v As Double, lambda As Double, rho As Double, nbrDec) As Double
Dim dich As Double, count As Integer, h As Double
Dim countDec


sscinv = 1
dich = ssc(sscinv, lambda, rho)

If (dich = v) Then
          Exit Function
End If

If (dich < v) Then
    h = 1
    count = 0
    countDec = -1
    While ((countDec < nbrDec) And (count < 2000))
        count = count + 1
        sscinv = sscinv + h
        If (ssc(sscinv, lambda, rho) = v) Then
                  Exit Function
        End If
        
        If (ssc(sscinv, lambda, rho) > v) Then
            countDec = countDec + 1
            sscinv = sscinv - h
            h = h / 10
        End If
    Wend
End If

If (dich > v) Then
    h = 0.1
    count = 0
    countDec = 0
    While ((countDec < nbrDec) And (count < 2000))
        count = count + 1
        sscinv = sscinv - h
        If (ssc(sscinv, lambda, rho) = v) Then
                Exit Function
        End If
        
        If (ssc(sscinv, lambda, rho) < v) Then
            countDec = countDec + 1
            sscinv = sscinv + h
            h = h / 10
        End If
       
    Wend
End If
End Function
\end{lstlisting}

\noindent \textbf{VB6 Subroutine for cumputing the \textit{cdf} $F$}. \label{ssc_cdf}
\begin{lstlisting}

Private Function ssc(x As Double, lambda As Double, rho As Double) As Double
Dim pi As Double

pi = 4 * Atn(1)
ssc = (1 / 2) + (1 / (2 * pi)) * (x - (lambda * (Cos(x) + 1)) + rho * Sin(x) 
      - (((lambda * rho) / 4) * (Cos(2 * x) - 1)))

End Function
\end{lstlisting}

\noindent \textbf{rssc : VB6 Subroutine for generating samples from the \textbf{ssc} distribution, which the sample in the file \textit{ssc100}}. \label{ssc_random}

\begin{lstlisting}
Sub rssc(tail As Integer, lambda As Double, rho As Double, nbrDec As Integer)
Dim ii As Integer, cheminD As String

cheminD = "G:\backup\DataGslo\gslo\TveSimulation\ssc\"
Open cheminD & "ssc100.txt" For Output As #1

Randomize Timer
For ii = 1 To tail
Print #1, sscinv(Rnd(1), lambda, rho, nbrDec)
Next
Close #1
End Sub
\end{lstlisting}

\end{document}